\documentclass[12pt]{article}
\usepackage[cp866]{inputenc}
\usepackage{amssymb}
\usepackage{graphics}

\newcommand\be{\begin{eqnarray}}
\newcommand\en{\end{eqnarray}}
\newcommand\ep{ $  e^+ e^- $ }
\newcommand\n{\nu, \tilde\nu}

\inputencoding{cp866}

\title{Efficiency of Electron-Positron Pair Production
by Neutrino Flux from Accretion Disk of a Kerr Black Hole}

\author{A.A.~Gvozdev{\footnote{e-mail: gvozdev@univ.uniyar.ac.ru}},
I.S.~Ognev{\footnote{e-mail: ognev@univ.uniyar.ac.ru}} \\[2mm]
{\it Yaroslavl State University}\\[2mm]
{\it 150000 Yaroslavl, Russia}}

\begin{document}

\maketitle

\begin{abstract}

{\normalsize
Dominant processes of neutrino production and neutrino-induced \ep-pair
production are examined in the model of a disk hyper-accreting onto a Kerr
black hole. The efficiency of plasma production by a neutrino flux from the disk, 
obtained for the both cases of presence and absence of a magnetic
field, is found to be no more than several tenths of percent and,
therefore, not enough for the origin of cosmological gamma-ray bursts.}
\end{abstract}

The origin of the gamma-ray burst is among the most important problems
to be solved. Various observations are in good agreement with a
phenomenological model implying that gamma-ray bursts are produced by an
ultrarelativistic \ep-plasma jet (fireball)~\cite{1}.
Observations indicate that gamma-ray bursts vary rapidly and some of
them arrive from cosmological distances. This makes to suggest that
the fireball is produced in a compact region and has a
huge energy of $ {\cal E} \gtrsim 10^{51} $~erg~\cite{15}.
One of the natural sources of such a fireball
could be neutrinos being able to carry away up to ten percent of the
gravitational energy released in a collapse in compact systems. Taking into
account the smallness of weak-interaction cross sections one can expect that
only a small fraction of the energy released is transferred to the
\ep-plasma, what in fact is enough to produce the fireball with
the energy pointed. However, the plasma produced can go out and remain the 
ultrarelativistic one (what is necessary for its further transformation into the 
observed gamma-ray burst) in a region with a sufficiently low baryon 
density~\cite{5}  

The conditions discussed can be realized in systems involving an accretion
disk around a Kerr black hole, e.g., failed supernova~\cite{2},
collapsar with hyper-accretion~\cite{3}, and hypernova~\cite{4}.
Due to the high accretion velocities and viscosity,
the density and temperature of the inner part of the disk can be
as high as $\rho \sim 10^{10} - 10^{11}$ g/cm$^3$ and $T \sim 5-10$ MeV,
so that neutrino luminosity reaches the value $L_\nu \sim 10^{53}$~erg/s.
At the same time, a region of a low baryon density can be formed in the
vicinity of the rotation axis~\cite{2,3}. Thus, a large neutrino flux
from the disk generates the plasma which can go out with the energy
sufficient for producing a gamma-ray burst.

It is important to note that strong magnetic fields can exist in the accretion
disk. The field strength in the viscous disk of the densities we are
interested in can reach values~\cite{12}:
\be
B \lesssim 10^{15} G
\left(
\frac{\alpha}{0.1}
\right)^{1/2}
\left(
\frac {c_s} {10^{9} cm / s }
\right)
\left(
\frac {\rho} {10^{11} g / cm^3 }
\right)^{1/2}  ,
\label{B}
\en
where $\alpha$ is the dimensionless viscosity parameter and
$c_s$ is the speed of the sound. The magnetic field can have a rather
complicated structure, however, for the processes considered only
the field strength is important.

The main process to create plasma by the neutrino flux in a rarefied
medium is considered as:
\be
\nu_i + \tilde\nu_i \Longrightarrow  e^+ + e^-
\;\;\; (i = e, \mu, \tau) ,
\label{Nue}
\en
It has been studied to apply to various astrophysical processes.
In early works, its influence on the explosion dynamics of a type-II
supernova was examined. For this purpose its luminosity in \ep-pairs
in the simplest models of neutrino blackbody emission
to the vacuum~\cite{6} and later on taking into account the fact that
the neutrino flux goes through a partially transparent medium of the
shell~\cite{7}. As the evidences in favor of the cosmological origin of the 
gamma-ray burst were accumulated, the process~(\ref{Nue}) has been considered as  
a possible energy source of the fireball~\cite{8}. 
The detailed numerical calculations of the fireball production
were performed only in recent papers ~\cite{2,3,9}, however the
magnetic field influence on plasma production
was not taking into account in these papers. Let us note that such an
influence can be substantial one in the  strong field. Indeed, in
this case new reactions of the \ep-pair production:
\be
\nu_i \Longrightarrow \nu_i +  e^+ + e^-  ,
\label{Bnu}
\\
\tilde\nu_i \Longrightarrow  \tilde\nu_i + e^+ + e^-
\label{Banu}
\en
are not only opened kinematically but can dominate as well.
The importance of these processes as a possible energy source of a
cosmological gamma-ray burst was pointed out first in the
paper~\cite{10}.

In the present paper we use the model of a disk hyper-accreting onto
a Kerr black hole to estimate the efficiency of \ep-plasma production
in processes involving neutrinos.
The efficiency is defined as the ratio of the  \ep-pair luminosity
$ L_{e^+ e^-} $ (energy emitted per unit time) to the neutrino luminosity
$(L_\nu + L_{\tilde\nu})$ from the disk:
\be
\epsilon = L_{e^+ e^-} / L_{tot} , \;\;\
L_{tot} =  L_\nu + L_{\tilde\nu}  .
\en
This paper focuses on deriving analytical expressions for the efficiency
of plasma production in the dominant neutrino processes
in a simplified model of accretion disk taking into account
a strong magnetic field. It's natural to expect that the efficiency will be 
estimated only. However such an approach would allow to show how the luminosity  
depends on the system parameters. We also neglect general relativity effects on 
plasma production. It's known, that the gravitational field can influence in two 
different ways: the neutrino redshift reduces the \ep-pair luminosity, where as 
the bending of neutrino trajectories increases it due to an increase in the 
collision frequency. An analysis of these effects indicates that the bending 
effect dominates only at sufficiently large radii of the last Keplerian orbit
($ R_0 \gtrsim 3 r_g $, where $r_g$ is the gravitational radius of a black
hole)~\cite{11}. However, the efficiency even in this case increases no more
than by a factor of 2.

As it was mentioned above, to ensure the required neutrino luminosity,
the inner part of the disk should have high densities and temperatures.
Such parameters can be attained in hyper-accretion onto a Kerr black
hole~\cite{3}. Leaving aside the question of stability of a system with such 
accretion rates, we use the results of~\cite{3} to determine the disk parameters. 
The gradients of density and temperature pointed in~\cite{3} allow us to
consider the neutrino-emitting part of the disk is uniform.
At an accretion rate of $ \dot M \sim 0.1 M_\odot /$s, the typical
densities and temperatures are $\rho \sim 10^{11}$~g/cm$^3$ and
$T \sim 5$~MeV, respectively. For these parameters, neutrinos are
predominantly emitted by Urca processes
\be
p + e^- \Longrightarrow n + \nu_e ,
\label{1} \\
n + e^+ \Longrightarrow p + \tilde\nu_e  .
\label{2}
\en
An analysis indicates that a magnetic field of $ B \sim 10^{15} $~G
has negligible effect on the cross sections for these reactions.
The neutrino mean free path in such a medium is estimated as
\be
l_\nu \sim 10 \; km
\left(
\frac{ 10^{11}  g / cm^3 }{ \rho }
\right)
\left(
\frac{ 5 MeV }{ T }
\right)^2 .
\label{lsrnu}
\en
Thus, the disk part under consideration can be treated as transparent to
neutrinos. Note that the typical times of establishing the
$\beta$-equilibrium in Urca processes (\ref{1}) and (\ref{2}) are
$ \tau_\beta \sim 10^{-2} $~s for the medium parameters used.
The characteristic dynamical accretion time can roughly be estimated
as the time it takes for a nucleon flux to pass through the
neutrino-emitting disk part and is also about $ \tau_d \sim 10^{-2} $~s.
Thus, the accreting matter does not arrive at the $\beta$-equilibrium
and, therefore, the parameter $ Y = N_p / (N_p + N_n) $ is indeterminate
and can vary in the interval
\be
Y_\beta < Y < 0.5   ,
\label{Y}
\en
where $N_p$ and $N_n$ are the proton and neutron number densities in the
disk and $Y_\beta$ is the $Y$ parameter at $\beta$-equilibrium
($ Y_\beta~\sim~0.1 $ for the densities and temperatures under
consideration).

Because the neutrino mean free path exceeds the characteristic
transverse dimension of the disk, neutrinos are free streaming
throughout the disk.
In this case, the neutrino luminosity is calculated in the standard way
by using the Lagrangian for the interaction of charged electron-neutrino
and nucleon currents in the low-energy approximation~\cite{13} and can be
represented as
\be
&&L_{\n}  = \int \omega F_{\n} d^3n ,
\label{Ln}
\\
&&F_{\n} = \frac { G_F^2 \cos^2 \theta_c ( 1 + 3 g_a^2 ) } { \pi }
\frac { \omega^2 N_{p,n} } { \exp [\omega / T \mp a ] + 1 }  .
\en
Here $\omega$ is the neutrino energy, $d^3n$ is the neutrino phase-space
element, $T$ is the temperature of the medium, 
$ a = (\mu - m_n + m_p) / T $, where $\mu$ is the electron chemical
potential, $m_n$ and $m_p$ are the neutron and proton masses, respectively, 
$N_n$ and $N_p$ are the neutron and proton number densities, respectively, 
$g_a$ is the axial constant of the charged nucleon current ($g_a \simeq$ 1.26
in the low-energy limit), $G_F$ is the Fermi constant and $\theta_c$ is the
Cabibbo angle. Integral~(\ref{Ln}) can easily be calculated for the
simplified model of a uniform disk. As a result, the (anti)neutrino
luminosity in the Urca processes from the disk is written as
\be
L_{\n} = \frac { (G_F \cos{\theta_c})^2 } { 2 \pi^3 }
( 1 + 3 g_a^2 ) N_{p,n} T^6 V I_5(\pm a)  ,
\;\;\;
I_s(a) = \int\limits_0^\infty \frac{y^s dy}{\exp{(y-a)} + 1},
\en
where $V$ is the emitting disk volume. The ratio of the neutrino luminosity
in the $Y$ interval~(\ref{Y}) is $ L_{\tilde\nu} / L_\nu << 1 $. As is seen
in the figure, even in the most favorable case of $\beta$-equilibrium, this
ratio is about one tenth and decreases very rapidly as $Y$ increases. Thus,
we set $ L_{tot} \simeq L_\nu $ in all cases unless this will lead to
confusion.

We calculate the \ep-pair luminosity for the case where the magnetic field
is sufficiently strong but the parameter $eB$ is much less than the
neutrino mean energy squared:
\be
m_e^2 \ll e B \ll \omega^2  ,
\label{wb}
\en
which is satisfied well in the case under consideration. Here, $m_e$ is the
electron mass and $e>0$ is the elementary charge. As was argued above,
it is most important to estimate the plasma production efficiency within a
small solid angle around the system rotation axis. Because the medium in
this cone is rarified, its effect on the processes can be neglected.
The magnetic field can have a complex structure in this region, but we
treat its field lines as directed along the rotation axis.

The electron-positron pair emissivity per unit volume in
reaction~(\ref{Nue}) is determined by
\be
Q_{\nu \tilde\nu \to e^+ e^-} = \int j \sigma  q_0 dN_\nu dN_{\tilde\nu} ,
\;\;\;
dN_{\n} = \frac { \omega^2 F_{\n} } { 8 \pi^3 R^2 }
dV  d \omega .
\label{q}
\en
where $\sigma$ is the cross section for the process,
$ j =  q^2 / ( 2 \omega_1 \omega_2 ) $ is the relative velocity in the
rest frame of one of the colliding particles,
$ dN_{\n} $ is the (anti)neutrino number density at a distance $R$ from
the 	element $dV$ of the isotropically emitting disk, and $ q = q_1 + q_2 $
is the 4-momentum transfer in the reaction. An analysis shows that the
magnetic field only slightly affects the cross section in the
approximation~(\ref{wb}):
\be
\sigma = \sigma_0
\bigg( 1 + O \Big( \frac{eB}{\omega^2} \Big) \bigg) ,
\;\;\;
\sigma_0 = \frac { G_F^2 } { 3 \pi }
\Big( c_v^2 + c_a^2 \Big) q^2  ,
\en
where $\sigma_0$ is the cross section for the process in vacuum, 
$ c_v = 1/2 + 2 \sin^2 \theta_W \simeq 0.96 $ and  $ c_a = 1 / 2 $ are,
respectively, the vector and axial constants of the charged
neutrino-electron current and $\theta_W$ is the Weinberg angle
($ \sin^2 \theta_W \simeq 0.23 $). Therefore, the luminosity in the
reaction under study can be estimated in the vacuum approximation. By
integrating Eq.~(\ref{q}) over the volume of a cone with a solid angle of
$ \Delta\Omega << 4 \pi $ along the black hole rotation axis, we obtain
the formula for the \ep-pair luminosity. It is reasonable to relate this
expression to the neutrino and antineutrino luminosity from the inner part
of the disk:
\be
L_{\nu \tilde\nu \to e^+ e^-} =
\frac { G_F^2 (c_v^2+c_a^2) } { 128 \pi }
L_\nu L_{\tilde\nu}
\frac {T}{R_0}
\left(
\frac {\Delta\Omega} {4 \pi}
\right)
\left[
\frac {I_6(a)} {I_5(a)}   +
\frac {I_6(-a)} {I_5(-a)}
\right] ,
\en
where $R_0$ is the radius of the last Keplerian orbit.

The luminosity for processes (\ref{Bnu}) and (\ref{Banu}) can be
calculated by using the expression obtained in~\cite{14} for the rate of
energy transfer to \ep-plasma per one neutrino. In approximation~(\ref{wb}),
this expression can be represented with logarithmic accuracy as
\be
\dot E = \frac { 7 G_F^2 ( c_v^2 + c_a^2 ) } { 432 \pi^3 }
( e B \omega \sin{\theta} )^2
\ln \bigg[
e B \omega \sin{\theta} / m_e^3
\bigg]   ,
\en
where $\theta$ is the angle between the initial neutrino momentum and the
magnetic field. By integrating this formula over neutrino distribution
$ dN_{\nu} $~(\ref{q}) and cone volume, we obtain the total luminosity
from the disk for the \ep-plasma produced in process~(\ref{Bnu}):
\be
L_{\nu \to \nu e^+ e^-} =
\frac { 7 G_F^2 ( c_v^2 + c_a^2) } { 1728 \pi^2 }
L_\nu (e B)^2 T R_0
\left(
\frac {\Delta\Omega} {4 \pi}
\right)
\frac { I_6(a) } { I_5(a) }
\ln \Big[ e B T / m_e^3 \Big]  .
\en
The antineutrino luminosity in process~(\ref{Banu}) is determined by the
same formula with replacement
\be
L_\nu \to L_{\tilde\nu},
\;\;\;
a \to -a  .
\en
Because the ratio $ L_{\tilde\nu} / L_\nu $ is small (see figure),
antineutrino reaction (\ref{Banu}) makes small contribution to
the total plasma luminosity.

Can new reactions~(\ref{Bnu}) and~(\ref{Banu}) be competitive with
basic process~(\ref{Nue}) of plasma production? The luminosity ratio for
these processes can be written as
\be
\frac{ L_{\nu \to \nu e^+ e^-} } { L_{\nu \tilde\nu \to e^+ e^-} }
= \eta
\left(
\frac{eB}{T^2}
\right)^2
\left(
\frac{l_{\tilde\nu}}{R_0}
\right)  ,
\en
where $\eta$ is a dimensionless constant of the order of unity. Therefore,
both processes can make comparable contributions to the \ep-pair luminosity
for the disk parameters used. However, the new processes become efficient
only if magnetic fields~(\ref{B}) attain their maximum strengths in the
disk--black hole system.

The efficiency of plasma production in process~(\ref{Nue}) is numerically
estimated as
\be
\epsilon_{\nu \tilde\nu \to e^+ e^-} \simeq  10^{-2}
\left(
\frac { L_{\tilde\nu} } { L_\nu }
\right)
\left(
\frac {\Delta\Omega} {4 \pi}
\right)
\left(
\frac { L_{tot} } { 10^{53} erg }
\right)
\left(
\frac {T} { 5 MeV }
\right)
\left(
\frac { 30  km } { R_0 }
\right) .
\label{eps}
\en
This expression depends strongly on the chemical composition of the
medium through the ratio $ L_{\tilde\nu} / L_{\nu} $ (see figure) and
decreases rapidly from its maximum value as the medium deviates from
the $\beta$-equilibrium. Thus, the efficiency of plasma production in the
absence of magnetic field does not exceed several tenths of percent
and becomes negligible at essential deviation of the nucleon medium
from $\beta$-equilibrium.

Similar estimation for process~(\ref{Bnu}) yields
\be
\epsilon_{\nu \to \nu e^+ e^-}
\simeq 2 \times 10^{-3}
\left(
\frac {\Delta\Omega} {4 \pi}
\right)
\left(
\frac { B } { 4 \times 10^{15} G }
\right)^2
\left(
\frac {T} { 5 MeV }
\right)
\left(
\frac { R_0 } { 30  km }
\right) .
\label{epsB}
\en
It is easy to see that the efficiency in this process is independent of the
disk chemical composition. This implies that the plasma production by a
single neutrino in a strong magnetic field may prevail over the
annihilation if there is a deviation from the $\beta$-equilibrium.
However, even in this case the efficiency of plasma production does not
exceed several tenths of percent and decreases quadratically as the
magnetic field decreases. Thus, the neutrino mechanism of plasma production
in collapsing systems with hyper-accretion is likely to be inefficient.
Ruffert and Janka~\cite{9} arrived at the similar conclusion for the model
of close binary system merging into a black hole.

We are grateful to N.V.~Mikheev and S.I.~Blinnikov for fruitful discussions
and to G.S.~Bisnovatyi-Kogan, N.I.~Shakura, M.E.~Prokhorov and
M.V.~Chistyakov for valuable comments. A.A.~G. thank to DESY theory group
for the warm hospitality and possibility to represent this work in
DESY Theory Workshop 2001 ''Gravity and Particle Physics''.

This work was supported in part by the Russian Foundation for Basic
Research (project no. 01-02-17334) and the Russian Ministry of
Education (project no. E00-11.0-5).

\newpage

\begin{figure}[htb]
\centerline{\includegraphics{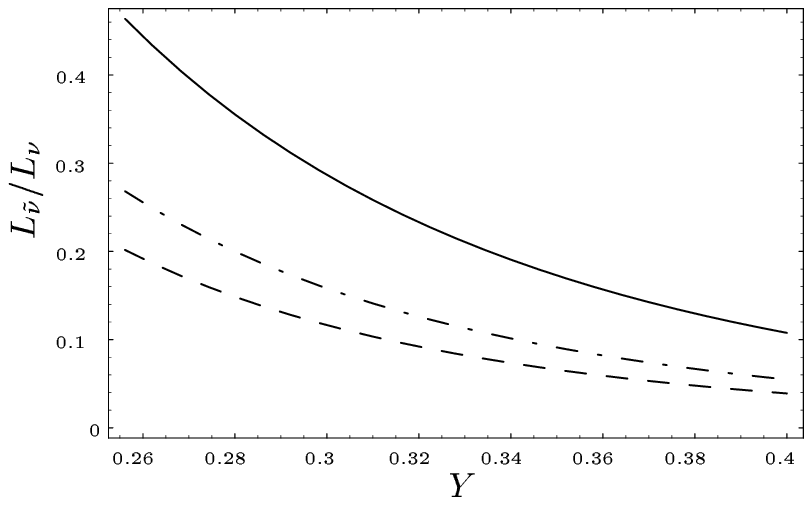}} 
\caption{
Neutrino luminosities ratio $L_{\tilde \nu} / {L_\nu}$ as function of parameter 
$Y$ with fixed $\rho$ and $T$. The solid line corresponds to 
$ \rho~=~10^{11}~g/cm^3,~T~=~7~MeV,~Y_\beta~=~0.30 $; 
the dash-dotted line corresponds to 
$ \rho~=~5~\times~10^{10}~g/cm^3,~T~=~5~MeV,~Y_\beta~=~0.26 $; 
dotted lines corresponds to 
$ \rho~=~10^{11}~g/cm^3, T~=~6~MeV, \; Y_\beta~=~0.24 $.
}
\end{figure}

\end{document}